\PassOptionsToPackage{table}{xcolor}

\documentclass[sigconf,screen]{acmart} 

\copyrightyear{2019} 
\acmYear{2019} 
\setcopyright{acmcopyright}
\acmConference[ESEC/FSE '19]{Proceedings of the 27th ACM Joint European Software Engineering Conference and Symposium on the Foundations of Software Engineering}{August 26--30, 2019}{Tallinn, Estonia}
\acmBooktitle{Proceedings of the 27th ACM Joint European Software Engineering Conference and Symposium on the Foundations of Software Engineering (ESEC/FSE '19), August 26--30, 2019, Tallinn, Estonia}
\acmPrice{15.00}
\acmDOI{10.1145/3338906.3340454}
\acmISBN{978-1-4503-5572-8/19/08}

\usepackage{graphicx}
\usepackage[toc,page]{appendix}
\usepackage{makecell}
\usepackage{multirow}
\usepackage{color, colortbl}
\usepackage{color}
\usepackage{ifxetex}
\definecolor{LightCyan}{rgb}{0.88,1,1}

\begin{document}

\title{An IR-based Approach Towards Automated Integration of Geo-spatial Datasets in Map-based Software Systems}

\author{Nima Miryeganeh}
\affiliation{%
  \institution{Department of Electrical and Computer Engineering \\
University of Calgary, Canada}
}
\email{seyednima.miryeganeh@ucalgary.ca}

\author{Mehdi Amoui}
\affiliation{%
  \institution{Localintel Inc.}
  \city{Calgary}
  \country{Canada}}
\email{mehdi@localintel.co}

\author{Hadi Hemmati}
\affiliation{%
  \institution{Department of Electrical and Computer Engineering \\
University of Calgary, Canada}
}
\email{hadi.hemmati@ucalgary.ca}

\renewcommand{\shortauthors}{Miryeganeh, Amoui and Hemmati}

\renewcommand{\shorttitle}{An IR-based Approach Towards Automated Integration of Geo-spatial Datasets ...}

\begin{abstract}
Data is arguably the most valuable asset of the modern world. In this era, the success of any data-intensive solution relies on the quality of data that drives it. Among vast amount of data that are captured, managed, and analyzed everyday, geospatial data are one of the most interesting class of data that hold geographical information of real-world phenomena and can be visualized as digital maps. Geo-spatial data is the source of many enterprise solutions that provide local information and insights. Companies often aggregate geospacial datasets from various sources in order to increase the quality of such solutions. However, a lack of a global standard model for geospatial datasets makes the task of merging and integrating datasets difficult and error prone. Traditionally, this aggregation was accomplished by domain experts manually validating the data integration process by merging new data sources and/or new versions of previous data against conflicts and other requirement violations. However, this manual approach is not scalable is a hinder toward rapid release when dealing with big datasets which change frequently. Thus more automated approaches with limited interaction with domain experts is required. As a first step to tackle this problem, we have leveraged Information Retrieval (IR) and geospatial search techniques to propose a systematic and automated conflict identification approach.  To evaluate our approach, we conduct a case study in which we measure the accuracy of our approach in several real-world scenarios and followed by interviews with Localintel Inc. software developers to get their feedbacks.
\end{abstract}




\maketitle

\section{Introduction}
% Paragraph 1: analytics + service +  Geospatial app
We generate, capture, and manage more data everyday. However, the real value of data is in further processing and analyzing it to gain insights. Data analytics software applications are class of software systems that are commonly used to explore and extract insights from multiple data sources. As local information is quite important, a popular subset of these systems cater geospatial data analytics for supporting spatial features of data records, along with other features, and commonly visualizing data and insights as maps.

% paragraph 4: frequently, conflict, costly manually
Since datasets from different sources may have different levels of abstraction, accuracy, and completeness the process of aggregating heterogeneous datasets can be challenging. In addition, some data sources may change frequently which out-dates the previous versions of the recorded data. The situation is even worse in case of geospatial data analytics, as identical geospatial entities from multiple sources can be captured and modeled differently.

As managing and organizing large number of geospatial datasets can be challenging, there are companies that offer Maps as a Service. The core offering of these companies is to gather data from multiple sources, clean, organize, manage, and update these datasets to provide their customers with up interactive maps that can represent several perspectives on-demand. As these service providers work with large number of datasets from multiple sources they inevitably encounter many conflicts while integrating new datasets or even when updating from an existing dataset. This is a significant scalability problem for companies in this domain, since the manual validation of frequent and large data integration phases are quite expensive. 
% Paragraph 5: Solution

Therefore, in this study, we propose a systematic validation approach, which semi-automatically aggregates geo-spatial datasets from heterogeneous sources, while keeping the system consistent after each merge. The idea is to reduce the number of potential conflicts that need a domain expert to validate the merge. Our approach consists of two phases:
\begin{itemize}
    \item Phase 1 -- \emph{Conflict Identification}: We propose a fully automated solution based on the similarity of spatial and non-spatial features of the merging data records. The spatial similarity functions are borrowed from Geographic Information System (GIS) research domain and  non-spatial similarities are taken from Information Retrieval (IR) domain. The novelty of this paper is combining these two similarities and evaluating them in an industrial context. 
    \item Phase 2 -- \emph{Conflict Resolution}: This semi-automated phase is our future work. We introduce its basic components in this paper, to show the full picture of the solution, but we do not report any results for phase 2.
\end{itemize}

The phase 1 results, based on our industrial case study, showed that we can identify the conflicts automatically, with over $95\%$ $Precision$ and $Recall$ values.  

\section{ Motivational Example } \label{ Localintel Case study }
Localintel is a software as a service company that provides market intelligence solutions to businesses that require such information to make informed decisions. Localintel aggregates data from various sources, including municipal, proprietary, and open-data to feed its web-based tools. These tools are basically web services that utilize map and dashboards to present local information and insights relevant to specific industries and their operating environment. 

\begin{figure}[!t]
\includegraphics[width=8.5cm]{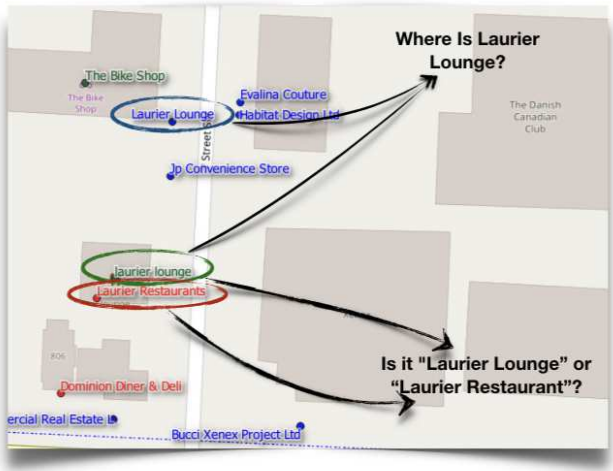}
\caption{Conflict - different colors represent different datasets.}
\label{ConflictsExamples}
\end{figure}

Most of Localintel tools contain multiple geospatial datasets from different sources. Each dataset has distinct properties, and quality of datasets varies. Depending on the data source, datasets are updated at different rates, on a regular basis or as needed. For example, to be able to build a layer which represents ``restaurants'' information as points on a map for downtown Calgary, Localintel receives datasets containing information about changes in restaurant and coffee shop businesses on a monthly basis. 

Given that input datasets are usually from different sources they can have different levels of accuracy in their spatial and non-spatial features. For example, Figure~\ref{ConflictsExamples} shows a sample case of integrating data from three different sources with inconsistent data about some points of interests. In this case, each dataset is showing a different location for an object called ``Laurier''. We also can not say if it is called a lounge or a restaurant. Such inconsistencies cause conflicts in the system and they need to be resolved before every merge of a new dataset to make sure that the system is always in a consistent state. 
Identifying and resolving these conflicts, manually, is only possible for few clients and small datasets. But it is definitely not scalable as the number of clients grows (with several layers being visualized for each) and datasets that are frequently being updated and merged into the system. Therefore, it is critical for the company to use an automated approach for data integration of geospatial datasets.

\section{The Semi-Automated Validation Approach}\label{The Semi-Automated Validation Approach}

\begin{figure}[!t] 
\includegraphics[width=8.5cm]{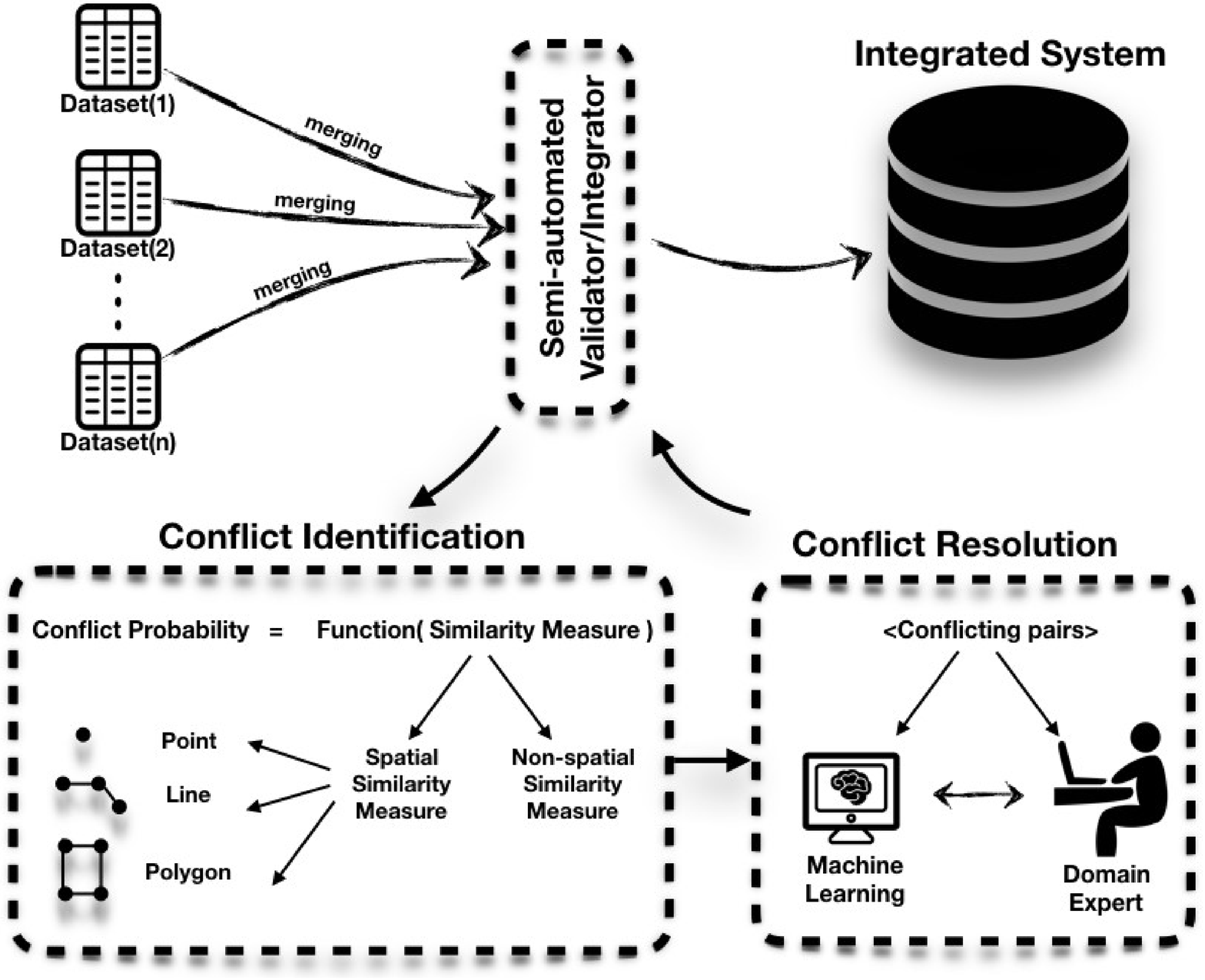}
\caption{The big picture of our semi-automated approach.}
\label{BigPicture}
\end{figure}

In our proposed approach, we aim to reduce the manual validation effort during data integration phase, either when merging new datasets or updating existing ones. Figure~\ref{BigPicture} summarizes the solution, which can be considered as a validation step before merging or updating data objects.

As illustrated in Figure~\ref{BigPicture}, the first step is to identify conflicting objects which introduce inconsistencies in the system. In the second step, we must decide on the resolving actions to remove the conflicts. For ease of reference, we denote the objects in the new dataset, which can be completely new or an updated version of an existing object, as $O_N$ and the existing objects of the system as $O_S$.

\subsection{Conflict Identification}
The sample case of Figure~\ref{ConflictsExamples} was an example of a case in which a data conflict is caused by objects corresponding to the same entity in the real world. At the first glance, it may seem that conflicting objects can be identified by getting the difference (``\textit{diff}'') of $O_N$ records and $O_S$ records. However, a simple \textit{diff} function is very sensitive and can only eliminate those $O_N$ records which there is an exact same $O_S$ for them and fails to detect the same objects with slight alterations. Therefore, to make sure that all the conflicts are resolved before a merge, we need a better solution than a simple \textit{diff} function. 

Assuming that there is no conflict between any two $O_N$ from one new dataset and all new datasets are internally consistent, and the current system is in a consistent state (there is no conflict between any two $O_S$) then each $O_N$ can potentially have conflicts with at most one  $O_S$ in the system. Hence, the conflict identification problem can be reformulated as \emph{searching for the most similar $O_S$ record per $O_N$}.

Now, in order to find out the similarity of a pair of objects ($<O_N,O_S>$), we need a similarity measure that takes both spatial and non-spatial features into account. Since these two types are completely different, we use two separate functions per feature type (spatial and non-spatial) and then combine the two measures to return one single similarity value, which can be considered as a ``conflict probability'' of the $<O_N,O_S>$ pair. 
The two similarity measures are defined as follow: 

\begin{itemize}
    \item \emph{Spatial Similarity Measure}: Geospatial objects are representations of entities abstracted as points, lines or polygons and stored as vectorized data. Therefore, to have a spatial similarity measure, we should be able to compare points, lines and polygons with each other.
   
    \item \emph{Non-Spatial (IR-based) Similarity Measure}: Non-spatial features can store a wide range of information about objects. They are usually textual or numerical attributes describing an object (e.g. name, address, telephone number, number of employees). So for non-spatial similarity, we adopt suitable textual or numerical similarity measures, from the IR domain, based on the context.  
\end{itemize}

% Textual and numerical similarity functions have been used in software engineering research and practiced in the past and there are many candidates that we can experiment with. However, the spatial features are less studied. Therefore, most of this paper focuses on the spatial similarity and how the overall measure can be made using them plus typical textual/numerical similarity functions. 

\subsubsection{Spatial Similarity Measure}
% A geospatial dataset is an abstraction of real world entities, which may include one or more dimensions. 
Different datasets can model the real world in different ways. 
For example, a dataset may abstract a river as a line while another dataset representing the same river as a polygon, or a building can be represented as a point or a polygon in different datasets. In addition to modeling/abstraction differences, there can also be differences between datasets caused by errors, such as slight alteration of a point location in two different datasets. Therefore, to capture all these different types of conflicting objects, we need to automate the following two class of comparisons:

\begin{itemize}
    \item CalssI: point2point; point2line; point2polygon
    % \item point2line comparison
    % \item  comparison
    \item CalssII: line2line; line2polygon; polygon2polygon
    % \item  comparison
    % \item  comparison
\end{itemize}

These two categories are based on the complexity of the involved objects and the similarity functions that we use for them.  Basically, the first class includes point objects and the second class involves more complicated shapes such as lines and polygons. 
\\

\textbf{Class I Comparison: }
In Class I comparisons, we are dealing with two objects, which at least one of them is a point. In such comparisons, we must measure the similarity of a point to other objects which are points or approximated as points. Note that for simplification purposes, when a point is being compared to a line or polygon, we compare the point with the centroid of the line or polygon. The similarity measure is symmetric, and we do not differentiate between point2line and line2point or point2polygon and polygon2point comparisons.

Searching among all objects in $O_S$ records to find similar cases to $O_N$ is quite expensive and does not pay off. Therefore,
in order to make the search space smaller, for each $O_N$, we search the adjacent objects by querying the database to retrieve the $O_S$ records located in (or overlapping with) a maximum distance of $\mathcal{E}$ from the reference object. This is a reasonable assumption, as conflicting objects are expected to be in close proximity. The next step is to assign a weight to all adjacent objects, which is achieved by the inverse of their (e.g., Euclidean) distance from the reference point. Finally, the normalized weights define the spatial conflict probabilities.
%In the next step, we divide each weight by the summation of all the adjacent objects' weight to make sure that we are capturing the similarities relatively. Finally, we normalize all the weights to transfer them in a range between 0 and 1, and we consider this as the spatial conflict probability of each pair.
\\

\textbf{Class II Comparison: }
In Class II comparisons, we are dealing with more complex shapes such as lines and polygons. Lines in vectorized data, are presented using a set of points with one start and one end point, while the start point $\neq$ the end point. Similarly, polygons are rendered on map using a set of points which define the border of that object. But unlike lines, the start point in a polygon must be equal to the end point, since polygons are closed shapes. But since both lines and polygons are demonstrated as set of points (edges), we do not differentiate between these two when adopting a similarity metric in class II comparisons.

For scalability concerns, as Class I, we reduce our search to only the adjacent objects.
So to find the search (neighboring) area for lines and polygons, again we consider circles with ratio of $\mathcal{E}$ around the vertices of their shapes. Then all the objects which are located within (or overlap with) this area are considered as adjacent objects. 
% Hence, we query the system database to retrieve all the objects which are located within (or overlap with) an area of $\mathcal{E}$ from the reference line/polygon.  
% \begin{table}[!t]
% \caption{Two classes of spatial comparisons.}
% \label{ComparisonClass}
% \centering
%     \resizebox{.8\hsize}{!}{%
%         \begin{tabular}{|c|| c | c | c |}
%             \hline
%              &\makecell{Point} & \makecell{Line} & \makecell{Polygon}\\
%             \hline
%             \makecell{Point} & \makecell{Class I} & \makecell{Class I}  &\makecell{Class I}\\
%             \hline
%             \makecell{Line} & \makecell{Class I} & \makecell{Class II} & \makecell{Class II}\\
%             \hline
%             \makecell{Polygon} &\makecell{Class I} & \makecell{Class II} & \makecell{Class II} \\
%             \hline
%         \end{tabular}%
%     }
% \end{table}
When we are dealing with points, defining a distance function is more obvious since there exist already many simple point2point distance functions such as Manhattan and Euclidean which we can adopt. But when we are dealing with more complex shapes such as lines and polygons, a suitable distance function might not be very intuitive. There exist several distance metrics for lines/polygons, which are mostly based on distance of vertices in two shapes \cite{shapeDistance}. Different measurements have been proposed to calculate the distance of two polygons/lines such as Hausdorff \cite{ModifiedHausdorff} and Chamfer, in which a distance is calculated using maximum and summation of distances of points in two shapes, respectively. Although Hausdorff and Chamfer can give us a good sense about the distance of two polygons/lines, but they are not the best in our context since they are very sensitive to addition or removal of edges/vertices. However, such a non-monotonic response to slight alterations is not desirable in our approach. On the other hand, the PoLiS \cite{Polic} metric, proposed by Avbelj et. al., is an alternative distance function, which is more promising and seems to be a better fit. So appropriate distance functions should be selected carefully base on the context of the features.   

\subsubsection{Non-Spatial Similarity Measure (IR-based)}
\label{TextualSim}
In the next step, we define a non-spatial conflict probability by calculating the textual/numerical distances between the two objects, using standard IR-based similarity functions. Since there are many different algorithms for textual similarity both for short textual segments \cite{shortTextSimilarity1,shortTextSimilarity2} and long textual documents \cite{longTextSimilarity}, we must make sure to adopt the best fit for each non-spatial feature. Therefore, we do not suggest a ``best'' similarity function for all features and rather recommend carefully experimenting to find and tune the best fit, per context.

In our study, to calculate the non-spatial similarity, we adopt Okapi BM25\cite{BM25} (which is implemented in Solr \cite{Solr}) to calculate the similarity score of documents. BM25 (BM stands for Best Matching) is a ranking function based on the probabilistic retrieval framework. BM25 and its variants can be considered as state-of-the-art TF-IDF-based functions, which are used in search engines as scoring algorithm to rank documents according to their relevance to a given search query. So each $O_S$($d$) is ranked based on the similarity score of its spatial and non-spatial fields to an $O_N$($q$), which is calculated as follows:

\begin{equation}
\label{eqn:01}
\fontsize{1}{10}\selectfont{
Score(q,d)=\sum_{t\in q}^{} {(tfNorm(t\in d).idf(t))}+DistBoost(q,d)}
$$\vspace{1mm}
\fontsize{9}{10}\selectfont{
$tfNorm = \frac{(freq * (k1 + 1))}{(freq + k1 * (1 - b + b * \frac{fieldLength)}{ avgFieldLength})}$\\
\vspace{2mm}
$idf(t)=\log(1 + \frac{(docCount - docFreq + 0.5)}{(docFreq + 0.5)}) $\\
\vspace{1mm}
$docCount$= Number of whole documents in the corpus\\
\vspace{1mm}
$docFreq$= Number of documents which include the term\\
\vspace{1mm}
$fieldLength$ = Number of terms in document(d) \\
\vspace{1mm}
$avgFieldLength$ = Average Number of terms in all documents\\
\vspace{1mm}
$K1$ = Term frequency normalization parameter\\
\vspace{1mm}
$b$ = Length normalization parameter \\
$DistBoost$: Calculated based on $\frac{1}{distance(q,d)}$, then normalized based on maximum non-spatial score }$$
\end{equation}
In equation \ref{eqn:01}, as mentioned, we use BM25 to calculate the non-spatial score of documents with default normalization values of K1=1.2 and b=0.75. Then to ensure that closer objects will eventually get a higher rank, we enhance this score by spatial similarity of objects (using inverse of distance of points) to tune the algorithm performance.

One important parameter that we have in this approach is the radius of search space ($\mathcal{E}$). A very large $\mathcal{E}$ can cause delays in the process of merging and makes it a very time consuming task, while choosing a very small $\mathcal{E}$ may result in failure in detection of some conflicts. Furthermore, to make sure that we are not flooding the domain experts with many conflict warnings consisting of many false positives, we define a threshold ($T$) on the conflict probability values to make the results more accurate. Note that ($T$) should be large enough to capture all the conflicts. 
These parameters must be carefully tuned based on the context and the domain expert's feedback, iteratively, to provide the best results. 

\section{Empirical Evaluation}
\label{CaseStudy}
To evaluate our approach, in real-world scenarios, we designed a set of experiments and interviews to assess the efficiency of our solution in different settings. In each experiment, we study a case from Localintel in which a new dataset merges into the existing database in the system. We also conducted a small set of interviews to get a feedback from the company developers on the approach. In the following subsections, we first describe the design of these experiments and interviews, then provide their results, and finally analyze the results to answer our research questions.   
\subsection{Objectives and Research Questions}
The goal of this section is to assess the effectiveness of our proposed automated approach for conflict identification, and compare it to a basic automated alternative.
The scope of the study is limited to datasets where each item, i.e. Point of Interest (PoI), is represented by a ``point''. We have broken down the goal into three research questions (RQs), as follows:

\begin{itemize}
    \item RQ1. How accurate is our proposed approach in terms of detecting existing conflicts compared to a baseline?
    
    \item RQ2. What is the effect of objects density and PoI type on the effectiveness of the results?
    
    \item RQ3. How effective and useful is our automated approach from practitioners point of view?
\end{itemize}   

\subsection{Study design}
The context of our study is a scenario at LocalIntel (our industry partner), where a new dataset is added to an existing database of PoIs. To answer our research questions, we have used three datasets that are the sources of data at LocalIntel, we always keep one as the existing database and merge a second one to the system. Unfortunately, due to confidentiality reasons, we can not name the datasets and will call them datasets A, B, and C.  

As mentioned, in all these experiments objects are stored as points and there is no line or polygon included in these datasets. For each new data point ($O_N$) there can be several possibilities, as follows: 
\begin{enumerate}
    \item Category 1: Non-conflicting objects (there is no corresponding $O_S$).
    \item Category 2: Conflicting objects:
    \begin{enumerate}
        \item There is an identical $O_S$.
        \item There is a corresponding $O_S$, but with practically ignorable differences.
        \item There is a corresponding $O_S$, but with practically significant differences.
    \end{enumerate}
\end{enumerate}

So as a first step toward a fully automated approach, we are seeking for an automated technique that a) can exclude category 1 objects from the others, and b) can detect conflicting objects in all three subgroups of category 2 and furthermore distinguishes between  category $1.c$ (the conflicts that need domain expert or machine learning for resolution) and the ones in category $1.a$ and $1.b$ (these are cases where one can automatically keep any of the two objects and delete the other to resolve the conflict).

\subsubsection{Implementation Details}

To implement our solution, we used Apache Solr\cite{Solr} which is an open-source tool for fast indexing and searching among documents. Using Solr, we indexed the system objects based on their location and textual features. In these experiments, we took ``business names'' as the only common feature to calculate the non-spatial similarity, but the tool can take any number of features to work with. 

We used StandardTokenizer\cite{Lucene} from Lucene to tokenize both documents and queries and after removing stopwords and transforming the tokens to their lowercase, using Porter stemmer, we unified all different forms that words can take. 

To combine spatial and non-spatial similarities, we first use the spatial similarity as a filter to narrow down the search space. This is done by taking a circle around each $O_N$, as its neighborhood area. Then for each $O_S$ in the neighborhood area of $O_N$, we calculate their textual similarity using the BM25 formula, which is described in Section\ref{TextualSim} which applies the distance boost to each adjacent objects score and take the $O_S$ with the highest score as the potential conflict for that $O_N$. $O_N$s with no $O_S$ passing the similarity threshold, are reported as non-conflicting objects. 

To tune the radius value for our experiments, we change the $\mathcal{E}$ value, from a search area of $\mathcal{E}$=100 meters to 250 meters (increments of 50m). We report the results of the tuning phase in the beginning of the results section.

\begin{table}[!t]
\caption{The details of experiments design}
\label{experimentDetails}
\centering
    \resizebox{1\hsize}{!}{%  
        \renewcommand{\arraystretch}{3}
        \begin{tabular}{|c|| c | c | c| c | c | c |}
            \hline
             \thead{\Large EXP\#}&\thead{\Large City} &\thead{\Large PoI} & \thead{\Large Dataset1} &
             \thead{\Large Dataset2} & \thead{\Large Area\\ ($m^2$)} & \thead{\Large Density\\ ($\frac{objects}{Km^2}$)}\\
            \hline\hline
            \makecell{EXP\#1} & Montreal & Restaurant & \makecell{B\\ (544 records)} & \makecell{A \\(1985 records)} & 32,834,195 & \makecell{$70-Low$}\\
            \hline
            \makecell{EXP\#2} & Seattle & Restaurant & \makecell{B \\(138 records)} & \makecell{C\\ (1442 records)} & 8,238,390  & \makecell{$190-Medium$}\\
            \hline
            \makecell{EXP\#3} & Calgary & Restaurant & \makecell{A\\(362 records)} & \makecell{C\\(616 records)} & 3,422,222 & \makecell{$280-High$}\\
            \hline
            \makecell{EXP\#4} & Calgary & Real Estate & \makecell{A\\(179 records)}& \makecell{C\\(274 records)} & 3,422,222 & \makecell{$130$}\\
            \hline
        \end{tabular}%
        \renewcommand{\arraystretch}{1}
      }
\end{table}
\subsubsection{RQ1 design}
To answer RQ1, we designed an experiment (Exp1), where we measure the effectiveness of our approach in detecting conflicts when merging two datasets (B is merged with A) of restaurants in downtown Montreal. The goal is to compare the results of our technique with a simpler alternative. 
The only related work that uses both spatial and non-spatial features during integration is the work by Seghal et. el.,  \cite{sehgal}. However, their textual similarity function is very naive (edit distance on the ``characters'' of textual features. So we improved this function and use a ``containment relation'' to serve as a baseline non-spatial similarity function. 

Therefore, In terms of implementation, both baseline and our proposed approach, follow the same routine: for each $O_N$, we search for the adjacent $O_S$ objects in the searching area and then for each pair we go through their common non-spatial features (which are previously converted to lowercase). 

However, with respect to non-spatial data, the similarity functions are different in baseline vs. our SDI approach. The baseline only checks for the ``containment relation'' between two texts, bidirectionally. If the containment relation is satisfied in all the common non-spatial features (the business name feature, in this study) in at least one direction, then the pair is reported as a conflict. Our proposed SDI approach, on the other hand, uses the explained TF-IDF-based approach as the non-spatial similarity function.

\subsubsection{RQ2 design}

To answer RQ2, we designed three new experiments (EXP 2,3 and 4). First, in Exp1 to 3, we analyze the performance of our proposed approach in cities with three different degrees of object density Montreal (``Low''), Seattle (``Medium''), and Calgary (``High''), for the restaurant datasets, and then we compare it to our baseline technique. Note that in these experimenets we always take two datsets out of three (A, B, and C), randomly, as existing vs. new dataset. Finally, in the next step (Exp4), we analyze the effect of PoI change on our solution effectiveness by taking a random city of the three studied ones (Calgary in this case) and change the PoIs from restaurant to Real Estate. 
Table\ref{experimentDetails} summarizes the features of all four experiments in RQ1 and RQ2.

\subsubsection{RQ3 design}
In order to validate our approach and assess its usefulness in practice, we arranged a small set of three interviews, in which we asked three software developers from Localintel's technical team, who did not know about this work prior to the interview, to work with our tool to merge a new dataset to Localintel system's database. The sessions took roughly an hour per developer where an interviewer (an author of the paper) helped them throughout the process. At the end of each session, the interviewer asked the developer a set of questions to get their high-level feedback on the approach and its results. The results is reported in the next section. 

\subsection{Study results}
In this section, we provide the results of the experiments and interviews to answer our three RQs. 
\begin{table}[!t]
\caption{Radius Tuning - The effect of $\mathcal{E}$ on the performance of solution in Exp1 (in terms of $Precision$ and $Recall$)}
\label{Tuning}
\centering
    \resizebox{1\hsize}{!}{%
        \begin{tabular}{|c|| c | c | c| c | c | c |}
            \hline
             \thead{Radius\\(meters)}&\thead{Total\\Number of\\ conflicts} &\thead{Conflicts\\correctly \\Detected} & \thead{Conflicts\\Wrongly\\Detected} &
             \thead{Missed\\conflicts} & \thead{Precision\\(\%)} & \thead{Recall\\(\%)}\\
            \hline
            100 & 254 & 242 & 0 & 12 & 100 & 95.27\\
            \hline
            150 & 254 & 243 & 0 & 11 & 100 & 95.66\\
            \hline
            \rowcolor{gray!30}
            200 & 254 & 245 & 0 & 9 & 100  & 96.45\\
            \hline
            250 & 254 & 243 & 2 & 11 & 99.18  & 95.66\\
            \hline
        \end{tabular}%
      }
\end{table}
\subsubsection{Tuning results}
As mentioned before, we first tune the search space radius ($\mathcal{E}$) to find the best search area for our experiments.  Table \ref{Tuning} summarizes the result of this analysis in terms of $Precision$ and $Recall$ for Exp1 (Restaurants of Montreal). 
As we can see, as we increase the $\mathcal{E}$ from 100 to 200 meters, the $Recall$ value increases 1.18 \% while the $Precision$ remains 100\%. However more increase in radius to 250 meters results in decrease in both $Precision$ and $Recall$. Based on these results and after consulting with the domain experts at Localintel, we take the $\mathcal{E} = 200 m$ as a default value for search area radius and we use this value in all the follow up experiments.

\subsubsection{RQ1 results}
Table \ref{Results} summarizes the performance of our solution for all the four experiments and compares it with the baseline method. 
To answer RQ1, we analyze Exp1 in which two datasets containing the restaurants of Montreal merge. In this experiment, there are 254 conflicts which are manually detected and labeled. We can observe that Baseline method can detect 184 conflicts, while missing the other 70 conflicts without producing any false positive. Although this is good in terms of $Precision$, but since this method fails to detect a big portion of conflicts, it is not very good in terms of $Recall$. However, our solution can detect 245 conflicts without generating any false positive, which increases the $Recall$ value from 72\% to 96\% (almost 24\% improvement) while keeping the $Precision$ 100\%. 
Observing the results from Exp1, we can now answer RQ1 as: `` With a substantial improvement to baseline method, our approach can detect conflicts with a relatively high precision and recall.''

\subsubsection{RQ2 results}
For Exp2 (restaurants of Seattle), with object density ``Medium'', again the baseline method performs well in terms of $Precision$, but misses 19 conflicts out of 89 labeled conflicts. On the other hand our solution, with missing only 4 conflicts, improves the $Recall$ from 78\% to 95\% (17\% improvement) while keeping the $Precision$ still very high (with only 2\% loss). 

Increasing the density to ``High'' in Exp3, the baseline method fails to detect 14 conflicts out of 340, however, our solution with missing only one conflict, can detect almost all the conflicts and improves the $Recall$ value from 95\% to 99\% (4\% improvement), with a loss of 1.5\% in $Precision$. 

Finally, with changing the PoI type in Exp4 to ``real estate agencies'', we again observe the same pattern.  Among 124 conflicts that have been manually labeled in this merge, baseline method can detect 111 of them, while missing the other 13. However, our solution can detect 119 conflicts and improve the $Recall$ from 89.5\% to 96\% (6.5\% improvement) while losing 2.5\% $Precision$ (misidentifying 3 conflicts).

Based on the above results, we can answer RQ2 as: `` PoI type and Object density do not have a significant impact on the effectiveness of the approach and our approach consistently dominates the baseline results.'' 

\begin{table}[!t]
\caption{Results of Exp1 to 4 for RQ1 and RQ2.}
\label{Results}
\centering
    \resizebox{1\hsize}{!}{
        \rowcolors{2}{gray!30}{gray!10}
        \begin{tabular}{|c|| c | c | c | c| c | c | c |}
            \hline
             \thead{\large EXP\#}&\thead{\Large{Method}}&\thead{\large Total\\\large Number of\\\large conflicts} &\thead{\large Conflicts\\\large correctly \\\large Detected} & \thead{\large Conflicts\\\large Wrongly\\\large Detected} &
             \thead{\large Missed\\\large conflicts} & \thead{\large Precision\\(\%)} & \thead{\large Recall\\(\%)}\\\hline
             \cellcolor{white}& Our Approach &254 & 245 & 0 & 9 & 100 & 96.45\\
             \multirow{-2}{*}{\cellcolor{white}EXP1}&Baseline &254 & 184 & 0 & 70 & 100 & 72.44\\\hline
             \cellcolor{white}& Our Approach  & 124 & 119 & 3 & 5 & 97.54  & 95.96\\
             \multirow{-2}{*}{\cellcolor{white}EXP2}& Baseline & 124 & 111 & 0 & 13 & 100  & 89.51\\\hline                  
             \cellcolor{white}& Our Approach & 340 & 339 & 5 & 1 & 98.54  & 99.7\\
             \multirow{-2}{*}{\cellcolor{white}EXP3}& Baseline& 340 & 326 & 0 & 14 & 100  & 95.88\\\hline
             \cellcolor{white}& Our Approach & 124 & 119 & 3 & 5 & 97.54  & 95.96\\
             \multirow{-2}{*}{\cellcolor{white}EXP4}& Baseline & 124 & 111 & 0 & 13 & 100  & 89.51\\\hline
        \end{tabular}
    }
\end{table}  

\subsubsection{RQ3 results}
Table \ref{Interview} summarizes the interview questions, and the complete interview can be found in appendix \ref{AppandixInterview}. 
Here to answer RQ3, we report a brief highlight of answers and  summarize the comments in each question. 
We use DEV1-3 to anonymize the interviewee identities.

\textbf{Discussion on the interview answers:}
In the first question, the goal is to validate our solution once again to ensure that we are approaching the problem from the right angle. Answers show that all developers agree that not only the tool works fine for now, with the point objects, but also they think that this systematic approach has the potential to grow bigger to solve the Spatial Data Integration (SDI) problem as a whole. 

These developers are people who are specialized in computer science and specially geo-spatial software systems and they have been working with such datasets for a long time. So in question 2 and 3, in order to understand their perspective, we asked them what would they do if they were supposed to make a solution for SDI, or improve/extend the current solution. From their answers, first we learn that our solution is approaching the problem from a right angel and we are on the right track. Although DEV1 mentioned that having a structured schema in the first place can be investigated as a possible solution, but since our tool is aimed to work with any kind of spatial dataset schema-independently, we do not consider schema conversion complexities to be able to come up with a fast solution that is compatible with all kinds of spatial data. Also, as depicted in Fig  \ref{BigPicture}, this tool can be considered as a preprocessing or a validation step for any next level processing such as schema-unification. Additionally, we collected some valuable ideas such as application of machine learning and fine tuning in identification and resolution steps to make the solution fully automated and more accurate. 
Finally, as the best jury, we asked them to give us their general feedback about this approach. As we can infer from their answers, they all think this systematic approach can be a valuable solution in SDI in companies that need to grow fast and scale up their database, and can be a good starting point toward a more accurate and inclusive solution.

\subsection{Discussion on Conflict Resolution}
\label{confres}
As mentioned in Fig\ref{BigPicture}, our solution consists of two steps, and after detecting the conflicts, we need an automated mechanism to resolve them before merging with the system. 

Although in some cases we may be able to  automatically resolve conflicts, but in many other cases we need the intervention of a domain expert to take the resolving decisions based on the origin of the conflicting objects and the semantics of the overlaps. Although we do not implement this part of the solution in our study, but we illustrate our vision by breaking down the problem and describing the structure of our data driven solution. To this end, first we categorize the conflicts based on their causes: 
\begin{table}[!t]
\caption{Interview Questions}
\label{Interview}
\resizebox{\linewidth}{!}{%
\begin{tabular}{|c|c|}
\hline
Q1 & \begin{tabular}[c]{@{}c@{}}Do you think the solution can work for Localintel\\ integration problem? How accurate you think the tool is?\end{tabular} \\ \hline
Q2 & \begin{tabular}[c]{@{}c@{}}What would you do if you were supposed \\ to design a tool to do the merging?\end{tabular}                                \\ \hline
Q3 & Do you have any suggestion to extend/improve the tool?                                                                                         \\ \hline
Q4 & Please provide any other feedbacks (free format).                                                                                                      \\ \hline
\end{tabular}%
}
\end{table}

\begin{itemize}
    \item \textit{Errors}: Conflicts that are caused by error-full data.
    \item \textit{Updates}: Conflicts that are caused by the outdated data.
    \item \textit{Abstraction}: Conflicts that are caused by objects with different levels of abstractions. 
    \item \textit{Completeness}: Conflicts that are caused by objects with different levels of completeness.
    \item \textit{False Positives}: Wrongly detected conflicts. 
\end{itemize}

Based on the cause of detected conflicts, there might be different resolving decisions taken by domain experts, i.e., a) keeping both of the objects, b) keeping one of them and deleting the other one, and c) making a new combined object. 
% Table \ref{CausesDecisions} summarizes different decisions per cause.

For example, conflicts due to Errors and Updates will be simply resolved by deleting the wrong or outdated object, or in conflicts which are due to completeness, we can combine two objects to make a new object that represents both, or in case of the abstraction conflicts, if one object can replace the other completely (it is more comprehensive and detailed) then the resolution is similar to the Update conflict case. 

Although in some cases we cannot resolve conflicts fully automated, and need for an expert supervision is inevitable, but in some other cases, with adoption of a suitable heuristic and machine learning algorithm, we can make the resolution process automated. The high-level idea here is that we keep recording the conflict resolution actions taken by the domain experts, per objects and datasets. Then when the training set is rich enough for classification, we can predict a resolving action automatically for a new identified conflict. Note that the machine-learning-based resolution is in our future work and has not been implemented/evaluated in this study. 

\section{Limitations and threats to validity}
In terms of the limitation of the current work, although our approach is showing promising results for the analyzed cases, but we still can not generalize it to all objects including lines and polygons. So for lines and polygons a suitable similarity metric should be carefully analyzed and defined in order to cover all class I and class II comparisons. Also, in these experiments, to calculate non-spatial similarity, we only used business name, however we can take more features into account depending on the involved datasets to improve the results.

In terms of the validity threats, we had a limited set of experiments and not exhaustively examined all combinations of the densities, PoI types, datasets, etc. Our interview study was also very small scale, more to get feedbacks rather than providing solid evidence of feasibilty or correctness. Therefore, we can not make any claims about the generalizabity of the study, before replicating this on more datasets and configurations.

We have done a simple tuning of radius on Exp1 and reused that for all experiments which might not be the optimal configurations, and thus a threat to the conclusion validity. There are also several other parameters in our algorithm that we have not tuned. However, a better tuning would potentially improve the result even more. To get a stronger conclusion we should also retry the experiments with several random datasets and PoIs and perform a proper set of statistical significant tests, which is in our future plan for replication.  

\section{Related Works}\label{Related Works}
Although SDI is not a new research topic and it is investigated by other researchers before, but it still lacks a robust, fast and scalable solution that can be applied in practice. While the spatial datasets are being gathered from heterogeneous sources continuously, the goal of SDI solutions is to make a united, integrated and single point access system that reflects the real world thoroughly and accurately \cite{related1,related2,related3}.
While some papers propose an integrated system for different use cases such as  river\cite{rivers} systems or bank's ATMs\cite{ATMs} datasets, geo-spatial datasets still require a multipurpose integrated system which can accurately represent reality from different aspects. 
Although having a general united data model \cite{DataModeling} can help us in having an integrated system, but with the emerge of noSQL databases, which offer more scalability and flexibility \cite{sqlVsNoSQL}, the urge for united data models is dimmer. But this does not imply that SDI challenges are solved. 
% Requirement checking and data validation still remain a challenging section in integration process of geospatial datasets.
Therefore, in this study, instead of focusing on an integrated model for geo-spatial datasets, we leverage IR to propose a solution which can integrate heterogeneous datasets. 

In this line of research, Beeri et al. \cite{beeri} have proposed two joint approaches, namely, the sequential and the holistic approaches for merging two or more datasets.  In another work, Safra et.al., \cite{POint2Point} propose an approach to find a location-based similarity of objects, which is based on one-sided nearest-neighbor join \cite{one-sided} of point objects. However these two solutions do not consider non-spatial features of data and just take the spatial features into account.

In another study, Sehgal proposes an ``entity resolution'' \cite{sehgal} method in which entities are matched using their non-spatial features and coordinates. However this solution is very naive when it comes to text similarity, since they use an edit distance function on the ``characters'' of each textual feature. 

Therefore, lack of a thorough and systematic approach which covers the whole problem entirely, inspired us to conduct this study in which we carefully investigate details of such systems and propose a systematic solution for SDI.

\section{Conclusion and Future Work}\label{conclusion}
Every day, terabytes of geo-spatial data are gathered from different sources and stored digitally as different datasets. 
A major challenge of aggregating heterogeneous datasets in one integrated system is that with each merge, we should make sure that the system remains in a consistent state. However, corresponding objects can sometimes cause conflicts in a system and entangle the process of merging. The current practice of identifying and resolving conflicts is very manual and thus expensive and not scalable.
In this study, which is a joint project in collaboration with industry, we introduce a new concept in SDI in which datasets can be continuously merged with the system database without requiring their schema to be converted to a unified model. With this concept a big portion of SDI effort which is related to schema uniforming is reduced to only data validation before inserting new records. In this paper, we tried to structure the problem and break it down to two steps of conflicts identification and resolution. Our initial evaluations showed a promising results both in terms of high accuracy of identification and consistency.

As our future work, we plan to conduct larger experiments with cities and datasets of different sizes to certify the efficiency of the solution, involve more complex objects (lines and polygons), and investigate machine-learning-based solution for the ``conflict resolution'' phase.  

% 
% If your work has an appendix, this is the place to put it.
\appendix

\section{Appendix: Interview Data}
\label{AppandixInterview}

This section is aimed to present results of the interviews which were conducted in our research. 
The participants in these interviews are from Localintel's technical team. Due to privacy concerns, we do not disclose their names and refer to them as Dev1, Dev2 and Dev3.

In each interview, after a one-hour session of  working with the tool in which an interviewer (an author of the paper) illustrates the instruction, each interviewee is asked a set of questions. 
Below is the summary of these interviews: 
\smallskip

\textbf{Q1. Do you think the solution can work for Localintel integration problem? How accurate you think the tool is?}
\smallskip 

\textbf{DEV1}: 
 \textit{``The solution created for conflict detection is very valuable in the data integration process. The ability to calculate a score on each conflict allows for continuous fine tuning on both the detection and resolution of conflicts. Although the accuracy of the conflict detection is not perfect, nor will it ever be, it's accuracy is high enough that it will be useful.''}

\textbf{DEV2}: 
\textit{``This is the first step in creating a single database that can be updated over time. We need to first resolve the conflicts between any new collections and our current collection before they can be merged.
\\
The tool has the potential to be very accurate with some tweaking of parameters. In its current state, it relies on a spatial search radius and a set of textual properties to assign a score which I think is very good.''}

\textbf{DEV3}: 
 \textit{``The solution has potential and could work towards helping solve Localintel's integration problem. The tool is fairly accurate. However, optimizations can be made to improve accuracy.''}
\smallskip 

\textbf{Q2. What would you do if you were supposed to design a tool to do the merging?}
\smallskip 

\textbf{DEV1}: 
 \textit{``If I were to build a similar product, I would have started by forcing some of the data into a known format, or schema, beforehand. Since the end goal is to fit all the data into an existing, well structured database, the process of identifying valuable fields and morphing them into a similar format would help greatly throughout the process of conflict detection and resolution.
 \\
Also, I would have started with a generic approach which could be used in a larger variety of use cases. Although less accurate from the start, being able to identify a broader spectrum of problems allows for the design of more robust tools which could be reused.''}

\textbf{DEV2}: 
 \textit{``I would have a similar solution. We only have a geometry and a set of properties for each data point. This solution takes both into account and uses an enterprise search platform (Solr) to compare the points in the best way possible.''}

\textbf{DEV3}: 
 \textit{``I would approach the problem in a similar way but would work on getting some training data and using machine learning to improve accuracy.
\\
Another possible way could be to cluster the points [lat, long, frequency of word in name]. The idea would be that the clusters would be the location of the restaurant or the conflict.''}
\smallskip

\textbf{Q3. Do you have any suggestion to extend/improve the tool? }
\smallskip 

\textbf{DEV1}: 
 \textit{``In order to improve the accuracy there should be a more automated way to test it. This could be done by creating a dataset, or several, in which you know the exact results that should be output. Once you know what the results should be, you can begin to refine the algorithms to get the scores you want and we can adjust the thresholds of conflict resolutions.''}

\textbf{DEV2}:
 \textit{``A machine learning algorithm could be applied to either this stage or the conflict resolution stage to assign higher or lower scores based on data properties of the points. For example, if fast food restaurants are more than often false positives, the algorithm could give them lower scores or suggest that there is a higher chance that they are distinct.''}

\textbf{DEV3}: 
 \textit{``... I would be giving a higher score based on several other aspects such as:
 \\
Source of dataset: Certain datasets could be more reliable than other and could be given a higher score.
\\
Type of point: In this case different types of restaurants such as fast food, chain restaurants, local restaurants. There are more fast food restaurants and multiple can be in the same area so depending on how far the points are they might not be conflicts.
\\
Location: If a fast food restaurant, it is likely that there would be another one close by. For example Starbucks. However, if there is a Starbucks in a more secluded area it is more likely that there would only be one.''}
\smallskip 

\textbf{Q4. Please provide any other feedback (free format):}
\smallskip 

\textbf{DEV1}: 
 \textit{``I have no doubt that it will prove to be an invaluable tool in our workflow. This tool is going to help us scale up our data integration by giving us confidence that errors will be kept at a minimum level. I like the use of solr for solving a unique problem such as this. I believe the approach taken can be applied to many datasets of the same type as well as many types of datasets.''}

\textbf{DEV2}:
 \textit{``It will be necessary to eventually resolve the conflicts into one schema to be able to use the data in a live production environment. NoSQL is probably a better fit for conflict detection though.''}

\textbf{DEV3}:
\textit{``I am interested in how this would work with lines and polygons.
It would be also interesting to see if clustering the points would result in the same results (you could have n-clusters based on the number of datapoints in a dataset)''}

\section{Acknowledgments}
This work is partially supported by the Natural Sciences and Engineering Research Council of Canada [EGP/521872-2017].

\bibliographystyle{ACM-Reference-Format}
\bibliography{main}

\end{document}